\input harvmac.tex

\noblackbox

\mathchardef\varGamma="0100
\mathchardef\varDelta="0101
\mathchardef\varTheta="0102
\mathchardef\varLambda="0103
\mathchardef\varXi="0104
\mathchardef\varPi="0105
\mathchardef\varSigma="0106
\mathchardef\varUpsilon="0107
\mathchardef\varPhi="0108
\mathchardef\varPsi="0109
\mathchardef\varOmega="010A

\font\mbm = msbm10
\font\Scr=rsfs10
\def\bb#1{\hbox{\mbm #1}}

\def\scr#1{\hbox{\Scr #1}}

\def\Mt{{\kern1em\hbox{$\tilde{\kern-1em{\scr M}}$}}}

\font\sScr=rsfs7
\def\sscr#1{\hbox{\sScr #1}}

%========================================================================
%          MACROS FOR REFERENCES
%========================================================================

%{{Nucl.\ Phys.}\/ {B \bf #1} (#2) #3}

%%%%%%%%%%%%%%%%%%%%%%%%%%%%%%%%%%%%%%%%%%
%%%%%%%%%%%%%%%%%%%%%%%%%%%%%%%%%%%%%%%%%%

%% References go here

\nref\fff{
I.~Antoniadis, C.~P.~Bachas and C.~Kounnas, ``Four-Dimensional Superstrings,''
Nucl.\ Phys.\  B {\bf 289} (1987) 87\semi
%%CITATION = NUPHA,B289,87;%%
H.~Kawai, D.~C.~Lewellen and S.~H.~H.~Tye,
``Construction of Fermionic String Models in Four-Dimensions,''
Nucl.\ Phys.\  B {\bf 288} (1987) 1.
%%CITATION = NUPHA,B288,1;%%
}

\nref\fsufive{
I.~Antoniadis, J.~R.~Ellis, J.~S.~Hagelin and D.~V.~Nanopoulos,
``The Flipped $SU(5) \times U(1)$ String Model Revamped,''
Phys.\ Lett.\  B {\bf 231} (1989) 65.
%%CITATION = PHLTA,B231,65;%%
}

\nref\slm{
A.~E.~Faraggi, D.~V.~Nanopoulos and K.~j.~Yuan,
``A Standard Like Model in the 4D Free Fermionic String Formulation,''
Nucl.\ Phys.\  B {\bf 335} (1990) 347\semi
%%CITATION = NUPHA,B335,347;%%
A.~E.~Faraggi,
``A New standard-like model in the four-dimensional free fermionic string
formulation,''
Phys.\ Lett.\  B {\bf 278} (1992) 131;
%%CITATION = PHLTA,B278,131;%%
``Construction of realistic standard-like models in the free fermionic
superstring formulation,''
Nucl.\ Phys.\  B {\bf 387} (1992) 239
[arXiv:hep-th/9208024]\semi
%%CITATION = NUPHA,B387,239;%%
G.~B.~Cleaver, A.~E.~Faraggi and D.~V.~Nanopoulos,
``String derived MSSM and M-theory Unification,''
Phys.\ Lett.\  B {\bf 455} (1999) 135
[arXiv:hep-ph/9811427].
  %%CITATION = PHLTA,B455,135;%%
}

\nref\alr{
I.~Antoniadis, G.~K.~Leontaris and J.~Rizos,
``A Three generation $SU(4) \times O(4)$ string model,''
Phys.\ Lett.\  B {\bf 245} (1990) 161\semi
%%CITATION = PHLTA,B245,161;%%
G.~K.~Leontaris and J.~Rizos,
``$N=1$ supersymmetric $SU(4)\times SU(2)_L \times SU(2)_R$ effective theory
from the weakly
coupled heterotic superstring,''
Nucl.\ Phys.\  B {\bf 554} (1999) 3
[arXiv:hep-th/9901098].
%%CITATION = NUPHA,B554,3;%%
}

\nref\cfs{
G.~B.~Cleaver, A.~E.~Faraggi and C.~Savage,
``Left-right symmetric heterotic-string derived models,''
Phys.\ Rev.\  D {\bf 63} (2001) 066001
[arXiv:hep-ph/0006331].
%%CITATION = PHRVA,D63,066001;%%
}

\nref\review{
For a recent and older review see {\it e.g.}:
J.~R.~Ellis, A.~E.~Faraggi and D.~V.~Nanopoulos,
``M-theory model-building and proton stability,''
Phys.\ Lett.\  B {\bf 419} (1998) 123
[arXiv:hep-th/9709049]\semi
%%CITATION = PHLTA,B419,123;%%
A.~E.~Faraggi,
``Phenomenological survey of free fermionic heterotic-string models,''
arXiv:0809.2641 [hep-th]
%%CITATION = ARXIV:0809.2641;%%
and references therein.
}

\nref\nilles{ See {\it e.g.}
H.~P.~Nilles, S.~Ramos-Sanchez, M.~Ratz and P.~K.~S.~Vaudrevange,
``From strings to the MSSM,''
Eur.\ Phys.\ J.\  C {\bf 59} (2009) 249
[arXiv:0806.3905 [hep-th]]\semi
%%CITATION = EPHJA,C59,249;%%
M.~Blaszczyk, S.~G.~Nibbelink, M.~Ratz, F.~Ruehle, M.~Trapletti and
P.~K.~S.~Vaudrevange,
``A $\bb{Z}_2 \times \bb{Z}_2$ standard model,''
Phys.\ Lett.\  B {\bf 683} (2010) 340
[arXiv:0911.4905 [hep-th]].
 %%CITATION = PHLTA,B683,340;%%
}

\nref\ploger{
F.~Ploger, S.~Ramos-Sanchez, M.~Ratz and P.~K.~S.~Vaudrevange,
``Mirage Torsion,''
JHEP {\bf 0704} (2007) 063
[arXiv:hep-th/0702176].
%%CITATION = JHEPA,0704,063;%%
}

\nref\foc{
A.~E.~Faraggi,
``$\bb{Z}_2 \times \bb{Z}_2$ orbifold compactification as the origin of
realistic free fermionic models,''
Phys.\ Lett.\  B {\bf 326} (1994) 62
[arXiv:hep-ph/9311312]\semi
%%CITATION = PHLTA,B326,62;%%
P.~Berglund, J.~R.~Ellis, A.~E.~Faraggi, D.~V.~Nanopoulos and Z.~Qiu,
``Toward the M(F)-theory embedding of realistic free-fermion models,''
Phys.\ Lett.\  B {\bf 433} (1998) 269
[arXiv:hep-th/9803262],
%%CITATION = PHLTA,B433,269;%%
``Elevating the free-fermion $\bb{Z}_2 \times \bb{Z}_2$ orbifold model to a
compactification of F-theory,''
Int.\ J.\ Mod.\ Phys.\  A {\bf 15} (2000) 1345
[arXiv:hep-th/9812141].
%%CITATION = IMPAE,A15,1345;%%
}

\nref\parti{
A.~E.~Faraggi,
``Partition functions of NAHE-based free fermionic string models,''
Phys.\ Lett.\  B {\bf 544} (2002) 207
[arXiv:hep-th/0206165]\semi
%%CITATION = PHLTA,B544,207;%%
A.~E.~Faraggi and E.~Manno,
``Little heterotic strings,''
arXiv:0908.2034 [hep-th]\semi
%%CITATION = ARXIV:0908.2034;%%
A.~E.~Faraggi and M.~Tsulaia,
``Interpolations Among NAHE-based Supersymmetric and Nonsupersymmetric String
Vacua,''
Phys.\ Lett.\  B {\bf 683} (2010) 314
[arXiv:0911.5125 [hep-th]].
%%CITATION = PHLTA,B683,314;%%
}

\nref\fknr{
A.~E.~Faraggi, C.~Kounnas, S.~E.~M.~Nooij and J.~Rizos,
``Classification of the chiral $\bb{Z}_2 \times \bb{Z}_2$ fermionic models in
the  heterotic superstring,''
Nucl.\ Phys.\  B {\bf 695} (2004) 41
[arXiv:hep-th/0403058].
%%CITATION = NUPHA,B695,41;%%
}

\nref\fkr{
A.~E.~Faraggi, C.~Kounnas and J.~Rizos,
``Chiral family classification of fermionic $\bb{Z}_2 \times \bb{Z}_2$
heterotic orbifold models,''
Phys.\ Lett.\  B {\bf 648} (2007) 84
[arXiv:hep-th/0606144],
%%CITATION = PHLTA,B648,84;%%
``Spinor-vector duality in fermionic $\bb{Z}_2 \times \bb{Z}_2$ heterotic
orbifold models,''
Nucl.\ Phys.\  B {\bf 774} (2007) 208
[arXiv:hep-th/0611251].
%%CITATION = NUPHA,B774,208;%%
}

\nref\forste{
A.~E.~Faraggi, S.~Forste and C.~Timirgaziu,
``Z(2) x Z(2) heterotic orbifold models of non factorisable six dimensional
toroidal manifolds,''
JHEP {\bf 0608}, 057 (2006)
[arXiv:hep-th/0605117].
%%CITATION = JHEPA,0608,057;%%
}

\nref\emanno{
E.~Manno,
``Semi-realistic Heterotic $\bb{Z}_2 \times \bb{Z}_2$ Orbifold Models,''
arXiv:0908.3164 [hep-th].
%%CITATION = ARXIV:0908.3164;%%
}

\nref\spinvecduality{
A.~E.~Faraggi, C.~Kounnas and J.~Rizos,
``Spinor-Vector Duality in $N=2$ Heterotic String Vacua,''
Nucl.\ Phys.\  B {\bf 799} (2008) 19
[arXiv:0712.0747 [hep-th]]\semi
%%CITATION = NUPHA,B799,19;%%
T.~Catelin-Jullien, A.~E.~Faraggi, C.~Kounnas and J.~Rizos,
``Spinor-Vector Duality in Heterotic SUSY Vacua,''
Nucl.\ Phys.\  B {\bf 812} (2009) 103
[arXiv:0807.4084 [hep-th]].
%%CITATION = NUPHA,B812,103;%%
}

\nref\vafawitten{
C.~Vafa and E.~Witten,
``On orbifolds with discrete torsion,''
J.\ Geom.\ Phys.\  {\bf 15} (1995) 189
[arXiv:hep-th/9409188].
%%CITATION = JGPHE,15,189;%%
}

\nref\nahe{
A.~E.~Faraggi and D.~V.~Nanopoulos,
``Naturalness of three generations in free fermionic $\bb{Z}_2 \times \bb{Z}_4$
string models,''
Phys.\ Rev.\  D {\bf 48} (1993) 3288\semi
%%CITATION = PHRVA,D48,3288;%%
A.~E.~Faraggi,
``Toward the classification of the realistic free fermionic models,''
Int.\ J.\ Mod.\ Phys.\  A {\bf 14} (1999) 1663
[arXiv:hep-th/9708112].
%%CITATION = IMPAE,A14,1663;%%
}

\nref\narain{
K.~S.~Narain,
``New Heterotic String Theories In Uncompactified Dimensions $< 10$,''
Phys.\ Lett.\  B {\bf 169} (1986) 41\semi
%%CITATION = PHLTA,B169,41;%%
K.~S.~Narain, M.~H.~Sarmadi and E.~Witten,
``A Note on Toroidal Compactification of Heterotic String Theory,''
Nucl.\ Phys.\  B {\bf 279} (1987) 369.
%%CITATION = NUPHA,B279,369;%%
}

\nref\higgsmattersplit{
A.~E.~Faraggi,
``Higgs-matter splitting in quasi-realistic orbifold string GUTs,''
Eur.\ Phys.\ J.\  C {\bf 49} (2007) 803
[arXiv:hep-th/0507229].
%%CITATION = EPHJA,C49,803;%%
}

\nref\saul{
K.~S.~Choi, S.~Groot Nibbelink and M.~Trapletti,
``Heterotic SO(32) model building in four dimensions,''
JHEP {\bf 0412} (2004) 063
[arXiv:hep-th/0410232]\semi
%%CITATION = JHEPA,0412,063;%%
H.~P.~Nilles, S.~Ramos-Sanchez, P.~K.~S.~Vaudrevange and A.~Wingerter,
``Exploring the SO(32) heterotic string,''
JHEP {\bf 0604} (2006) 050
[arXiv:hep-th/0603086]\semi
%%CITATION = JHEPA,0604,050;%%
S.~Ramos-Sanchez,
``Towards Low Energy Physics from the Heterotic String,''
Fortsch.\ Phys.\  {\bf 10} (2009) 907
[arXiv:0812.3560 [hep-th]].
%%CITATION = FPYKA,10,907;%%
}

\nref\openstring{  M.~Bianchi and A.~Sagnotti, ``On the systematics of open
string theories,''
  Phys.\ Lett.\  B {\bf 247} (1990) 517
   %%CITATION = PHLTA,B247,517;%%
\semi
  C.~Angelantonj and A.~Sagnotti,
``Open strings,''
  Phys.\ Rept.\  {\bf 371} (2002) 1
  [Erratum-ibid.\  {\bf 376} (2003) 339]
  [arXiv:hep-th/0204089].
  %%CITATION = PRPLC,371,1;%%
}

\nref\orbifolds{
L.~J.~Dixon, J.~A.~Harvey, C.~Vafa and E.~Witten,
``Strings On Orbifolds,''
Nucl.\ Phys.\  B {\bf 261} (1985) 678,
%%CITATION = NUPHA,B261,678;%%
``Strings On Orbifolds. 2,''
Nucl.\ Phys.\  B {\bf 274} (1986) 285.
%%CITATION = NUPHA,B274,285;%%
}

\nref\gepner{
D.~Gepner,
``New Conformal Field Theories Associated with Lie Algebras and their Partition
Functions,''
Nucl.\ Phys.\  B {\bf 290}, 10 (1987),
%%CITATION = NUPHA,B290,10;%%
``Space-Time Supersymmetry in Compactified String Theory and Superconformal
Models,''
Nucl.\ Phys.\  B {\bf 296} (1988) 757.
%%CITATION = NUPHA,B296,757;%%
}

\nref\chang{
D.~Chang and A.~Kumar,
``Mechanisms of spontaneous symmetry breaking in the fermionic construction of
superstring models,''
Phys.\ Rev.\  D {\bf 38} (1988) 1893.
%%CITATION = PHRVA,D38,1893;%%
``Twisted Thirring interaction and gauge symmetry breaking in $N=1$
supersymmetric superstring models,''
Phys.\ Rev.\  D {\bf 38} (1988) 3734.
%%CITATION = PHRVA,D38,3734;%%
}

%%%%%%%%%%%%%%%%%%%%%%%%%%%%%%%%%%%%%%%%%%%%%%%%%

%:Title

\Title{\vbox{
\rightline{}
\rightline{DFTT 2010/4}
\rightline{LTH-868}
}}
{\vbox{
\centerline{Spinor-Vector Duality}
\bigskip
\centerline{in Heterotic String Orbifolds}
}}

\centerline{Carlo Angelantonj$^\dagger$,
Alon E. Faraggi$^\star$ and
Mirian Tsulaia$^\star$\footnote{$^1$}{Associate member of the Centre for
Particle Physics and Cosmology,
Ilia State University, 0162 Tbilisi, Georgia
}
}

\medskip
\centerline{\it
$^\dagger$ Dipartimento di Fisica Teorica and INFN Sezione di Torino}
\centerline{\it Via P. Giuria 1, I--10125 Torino}
\centerline{\it $^\star$ Department of Mathematical Sciences,
		University of Liverpool,}
\centerline{\it Liverpool L69 7ZL, United Kingdom}

\vskip 0.6in

\centerline{\bf Abstract}

\noindent
The three generation heterotic-string models in the free fermionic
formulation are among the most realistic string vacua constructed to
date, which motivated their detailed investigation. The classification of
free fermion heterotic string vacua has revealed a duality under the
exchange of spinor and vector representations of the SO(10) GUT
symmetry over the space of models. We demonstrate the existence of the
spinor-vector duality using orbifold techniques, and elaborate on the
relation of these vacua to free fermionic models.

\Date{March 2010}

%:Introduction

\newsec{Introduction}

String theory provides a self-consistent framework to describe quantum gravity
and particle
physics in a unified way. Several approaches to particle phenomenology have
been pursued, based on heterotic string compactifications, orientifold
constructions, M-theory compactification on manifold of special
holonomy and/or F-theory techniques. 
All these scenarios have brought new interesting ideas to particle physics and
string theory, though none can be considered as ``fully realistic''. Among the
various approaches, heterotic string theory still seems to be a preferred
candidate to build quasi-realistic models, and particularly promising is the
free-fermionic
construction of heterotic vacua \fff.
Although these constructions are typically formulated at special points in the
moduli space and thus lack an
apparent geometric description, over the last two decades they have shown to be
very powerful tools to develop phenomenological string vacua
\refs{\fsufive{--}\cfs}.
Three generation models with the correct Standard Model charge assignments,
as well as the canonical SO(10) embedding of the weak hypercharge have been
constructed,
and various phenomenological issues have been further explored \review.

More recently, classes of quasi-realistic heterotic string models have also
been constructed, based on orbifold techniques \refs{\nilles,\ploger}, that
also allow to explore the underlying moduli dependence of couplings and gauge
groups. It should be stressed, however, that the two formulations --- in terms
of free fermions or in
terms of free bosons --- are closely related and the corresponding string
vacua can be described equivalently using the two formalisms. Indeed,  the
free-fermionic constructions correspond in general to
${\bb Z}_2^n$ toroidal orbifolds, when the geometric data of the six-torus are
chosen to
correspond to special points of moduli space.

It is therefore necessary to develop a dictionary between the two languages, in
such a way to
be able to address questions related to moduli dynamics within a given
free-fermionic vacuum.
While the equivalence is anticipated, writing a detailed dictionary is often
non-trivial. A first attempt
to establish such a link was done in \refs{\foc,\parti} in the context of
${\scr N}=4$ toroidal compactification.
In fact, in many quasi-realistic free-fermionic constructions the starting
(four-dimensional) gauge symmetry is
${\rm SO} (16) \times {\rm SO} (16)$, rather than the more conventional ${\rm
E}_8 \times {\rm E}_8$ symmetry --- or, at times, the SO(32) symmetry ---
typically considered in bosonic constructions.
In  the free-fermionic realisations the two choices correspond to different
solutions of the modular invariance constraints. Then in terms of free bosons
the two choices can be shown to depend
on the possibility to turn on or off discrete torsion in certain freely acting
$\bb{Z}_2 \times \bb{Z}_2$ orbifolds \parti. Alternatively, this amounts to
different choices of Wilson lines and geometrical backgrounds.

Clearly, in order to build quasi-realistic chiral models, ${\scr N}=4$
supersymmetry ought to
be broken to ${\scr N}=1$, and eventually to ${\scr N} =0$. This can be
achieved by performing a geometric $\bb{Z}_2 \times \bb{Z}_2$ projection on the
${\scr N}=4$ vacua. Additionally, this projection breaks the ${\rm SO} (16)
\times {\rm SO} (16)$ gauge group to the more phenomenologically  appealing
${\rm SO} (10) \times {\rm U} (1)^3 \times {\rm SO} (16)$, while chiral matter
emerges from the twisted sectors.
Although in free-fermionic set-ups there are many consistent solutions with
different
low-energy chiral spectra \refs{\fknr,\fkr}, it seems that much fewer choices
are present
in the free bosonic case. However, this is in contrast with the expectation
that the two
formulations are equivalent. In particular, naively adding a $\bb{Z}_2$
geometric twist to the model
of ref. \parti\ retains the vectorial representations in the massless spectrum
rather than the spinorial ones \emanno.

In this paper we make a step forward in the direction of a better understanding
of
the connection between the formulation of the heterotic string in terms of free
bosons and free fermions.
A particular issue we would like to address is the recently proposed {\it
spinor-vector duality} in
heterotic-string vacua \spinvecduality, which was observed using the free
fermionic language.
This new duality relates vacua with spinorial and vectorial representations of
orthogonal gauge groups, and it has been shown to hold in ${\scr N}=2$ and
${\scr N}=1$ free-fermionic models.
It was also suggested that the spinor-vector duality can be thought of as being
an
extension of mirror symmetry \spinvecduality. Indeed, mirror symmetry implies a
change in
the topology of the compactification manifold, that flips the sign of its Euler
number.
Equivalently, spinor-vector duality can be thought of as another
topology-changing operation.

To date spinor-vector duality has not been studied in the orbifold language.
In this paper we study this issue by analysing the ${\rm E}_8\times {\rm E}_8$
heterotic string compactified on the orbifold $T^6/ \bb{Z}_2 \times
\bb{Z}_2^{\prime}\times \bb{Z}_2^{\prime\prime}$.
The three $\bb{Z}_2$ operations correspond to the two supersymmetry preserving
freely acting twists of ref. \parti, while $\bb{Z}_2^{\prime\prime}$ reflects
four
internal coordinates and breaks ${\scr N}=4$ to ${\scr N}=2$. The ${\rm E}_8$,
and SO(16), symmetries are reduced by the
$\bb{Z}_2$ twist to ${\rm E}_7\times {\rm SU} (2)$, and ${\rm SO}(12)\times
{\rm SO}(4)$, respectively.
In this case the spinor and vector representations are both in the $56$
representation of ${\rm E}_7$, that decomposes as $(32,1)+(12,2)$ under its
maximal ${\rm SO} (12)\times {\rm SU} (2)$ subgroup.
Here, the $32$, and $12$, are the spinorial, and vectorial, representations, of
SO(12), respectively.
As the twisted sectors of the geometrical $\bb{Z}_2\times \bb{Z}_2$ orbifolds
preserve ${\scr N}=2$ supersymmetry, it is sufficient to study the
spinor-vector duality at this level rather than in the ${\scr N}=1$ models,
which are then obtained with an  additional $\bb{Z}_2$ twist. The partition
function associated to this $\bb{Z}_2 \times \bb{Z}_2^{\prime}\times
\bb{Z}_2^{\prime\prime}$ has eight independent
orbits, that admit seven discrete torsions taking the values $\pm1$.
Different choices of such discrete torsions clearly yield different spectra
and, among those, there are some that retain the spinorial representation and
others that retain vectorial one. As a consequence, we note the existence of a
transformation that maps between the cases, which reproduces the spinor-vector
duality map observed in ref. \spinvecduality\ within free fermionic
construction. As in the free fermion case \spinvecduality, the spinor-vector
duality exists at the ${\scr N}=2$ level, which is
obtained with a single $\bb{Z}_2$ twist acting on the internal coordinates.
Actually, the heart of the spinor-vector splitting is in the choice of the
${\scr N}=4$ vacuum, where
${\rm E}_8\times {\rm E}_8$ is broken to ${\rm SO}(16)\times {\rm SO} (16)$.
The additional $\bb{Z}_2^{\prime\prime}$ twist then
selects either the spinorial or the vectorial representation of the resulting
gauge group, and  the spinor-vector duality map depends nontrivially on the
discrete torsions, as we find in this paper. This is in a sense analogue with
the mirror symmetry analysis of ref. \vafawitten, where the single discrete
torsion of the $\bb{Z}_2\times \bb{Z}_2$ geometrical orbifold flips the Hodge
numbers of the internal manifold.

Our paper is organised as follows: in Section 2 we review the construction of
``quasi-realistic'' free-fermionic vacua and discuss the emergence of the
recently discovered vector-spinor splitting as a freedom in the choice of
generalised GSO (GGSO) phases. We then discuss, in Section 3, equivalent
constructions based on $\bb{Z}_2^n$ orbifolds of free bosons and show
explicitly how the vector-spinor splitting is, in this context, a consequence
of the freedom of turning on or off different discrete torsions. Section 4
concludes with some comments, while in the appendix we list various
combinations of characters that play a role in the constructions presented in
Section 3.

%:free-fermion
\newsec{Spinor-vector splitting in free-fermionic models}

In this section we discuss the spinor-vector splitting in free-fermionic models
(see \review\ for a more detailed introduction). In the free-fermionic
formulation of the heterotic string in four dimensions all the world-sheet
degrees of freedom, required to cancel the conformal anomaly, are represented
in terms of free fermions
propagating on the string world sheet \fff. In the light-cone gauge, the
world-sheet degrees of freedom then consist of two transverse left-moving
fermions $\psi^\mu_{1,2}$, superpartners of the space-time left-moving bosonic
coordinates, together with additional 62 purely internal
Majorana-Weyl fermions. Eighteen of them are left-moving,
$$
\chi^{1,..,6}, \quad y^{1,...,6}, \quad \omega^{1,...,6} \,,
$$
while the remaining 44 are right-moving
$$
{\overline y}^{1,...,6}\,,
\quad {\overline \omega}^{1,...,6}\,,
\quad {\overline \psi}^{1,..,5}\,,
\quad {\overline \eta}^{1,2,3}\,,
\quad {\overline \phi}^{1,...,8}\,.
$$
Under parallel transport around a non-contractible loop on the toroidal
world-sheet the fermionic fields pick up a phase, $ f~\rightarrow~-{\rm
e}^{i\pi\alpha(f)}f~,~~\alpha(f)\in(-1,+1].$
Each set of specified phases for all world-sheet fermions, around all the
non-contractible
loops is called the spin structure of the model. Such spin structures
are usually given in the form of 64 dimensional boundary condition vectors,
with each entry specifying the phase of the corresponding
world-sheet fermion. The basis vectors are constrained by string consistency
requirements, and completely determine the vacuum structure of the model.
The physical spectrum is then obtained by applying suitable GGSO projections.

The boundary condition basis defining a typical ``realistic free fermionic
heterotic string model'' is  constructed in two stages. The first stage
consists of the NAHE set, which is a set of five boundary condition basis
vectors,  $\{ 1 ,S,b_1,b_2,b_3\}$ \nahe
$$
\eqalign{
S &= \{\psi^{1,2}, \chi^{1,...,6}\}\,,
\cr
b_1 &= \{ \psi^{1,2}\,,
\chi^{1,2}\,, y^{3,..,6}\,|\, {\overline y}^{3,..6}\,, {\overline
\psi}^{1,..,5,}\,, {\overline \eta}^{1} \}\,,
\cr
b_2 &= \{ \psi^{1,2}\,, \chi^{3,4}\,, y^{1,2}\,, \omega^{5,6}\,| \,
{\overline y}^{1,2}\,, {\overline \omega}^{5,6}\,, {\overline
\psi}^{1,..,5,}\,,
{\overline \eta}^{2} \}\,,
\cr
b_3 &= \{ \psi^{1,2}\,, \chi^{3,4}\,,  \omega^{1,..,4}\,|  \,
{\overline \omega}^{1,..,4}\,, {\overline  \psi}^{1,..,5,}\,,
{\overline \eta}^{3} \}\,,
\cr}
$$
where, for simplicity, only the fields with $\alpha (f)=1$ are explicitly
indicated, while those that are not listed have $\alpha (f) =0$. After imposing
the GSO projection, the gauge group is ${\rm SO} (10)\times {\rm SO}(6)^3\times
{\rm E}_8$, and the vacuum enjoys  ${\scr N}=1$ supersymmetry. The second stage
of the construction consists of adding to the
NAHE set three (or four) additional boundary condition basis vectors, typically
denoted by $\{\alpha,\beta,\gamma\}$.  These additional basis vectors reduce
the number of chiral generations to three, one from each of the sectors $b_1$,
$b_2$ and $b_3$, and simultaneously break the SO(10) GUT symmetry to one of its
subgroups  \refs{\fsufive, \slm, \alr, \cfs}.

The correspondence of the NAHE-based free fermionic models with the orbifold
construction is illustrated
by extending the NAHE set, $\{ 1,S,b_1,b_2,b_3\}$, by one additional boundary
condition basis vector \foc,
\eqn\vectorx{
\xi_1= \{ {\overline \psi}^{1,..,5}, {\overline \eta}^{1,2,3} \}\,.
}
In this way, the orbifold construction involves an internal lattice with
nontrivial background fields \narain. Indeed, the subset of basis vectors
\eqn\neqfourset{
\{ 1,S,\xi_1,\xi_2\}\,, \qquad \quad \xi_2 = 1+ b_1+b_2 +b_3
}
generates a toroidally compactified model with ${\scr N}=4$ space-time
supersymmetry and ${\rm SO}(12)\times {\rm E}_8\times {\rm E}_8$ gauge group.
Here the enhanced
${\rm U} (1)^6 \to {\rm SO}(12)$ gauge symmetry is precisely due to the choice
of the internal SO(12) lattice, with non trivial $B_{ij}$ and $G_{ij}$
backgrounds.
Adding the two basis vectors $b_1$ and $b_2$ to the set \neqfourset\
corresponds then to the $\bb{Z}_2\times \bb{Z}_2$ orbifold model with standard
embedding, and Hodge numbers $h_{11}=27$ and $h_{21}=3$.
We note that the Euler characteristic of this orbifold differs
from that of a $Z_2\times Z_2$ orbifold at a generic point
in the moduli space due to identification of fixed 
points by an internal lattice shift \refs{\foc,\parti,\forste}.

The effect of the additional basis vector $\xi_1$ of eq.
\vectorx, is to separate the gauge degrees of freedom, spanned by
the world-sheet fermions $\{{\bar\psi}^{1,\cdots,5},
{\bar\eta}^{1},{\bar\eta}^{2},{\bar\eta}^{3},{\bar\phi}^{1,\cdots,8}\}$,
from the internal compactified degrees of freedom $\{y,\omega\vert
{\bar y},{\bar\omega}\}^{1,\cdots,6}$. In this construction, one actually has
the freedom of flipping the sign of some GGSO phases, compatibly with modular
invariance. In particular, the choice
\eqn\changec{
c{\xi_1\choose \xi_2}\rightarrow -c{\xi_1\choose \xi_2}\,,
}
breaks the  ${\rm E}_8 \times {\rm E}_8$ gauge symmetry down to ${\rm SO}(16)
\times {\rm SO}(16)$, that is instrumental for getting the GUT gauge group
SO(10) since, after the inclusion of the vectors $b_1$ and $b_2$, ${\rm SO}(16)
\times {\rm SO}(16) \to {\rm SO}(10) \times {\rm U}(1)^3 \times {\rm SO}(16)$.

In the ``realistic free fermionic
models'' this is achieved by the vector $2\gamma$ \foc
\eqn\vectortwogamma{
2\gamma = \{ {\overline \psi}^{1,...,5}\,, {\overline \eta}^{1,2,3}\,,
{\overline \phi}^{1...,4} \} \,,
}
that has the same effect of breaking ${\rm E}_8 \times {\rm E}_8$ gauge
symmetry down to
${\rm SO}(16) \times {\rm SO}(16)$, and then to ${\rm SO}(10) \times {\rm
U}(1)^3 \times {\rm SO}(16)$
after the standard $\bb{Z}_2 \times \bb{Z}_2$ project is enforced.

The freedom in \changec\ actually corresponds to a discrete torsion. In fact,
at the level of the ${\scr N}=4$ Narain model generated by the set \neqfourset,
one can build two different vacua, ${\scr Z}_+$ and ${\scr Z}_-$, depending on
the sign
of the discrete torsion in eq. \changec. The first, say ${\scr Z}_+$,
produces the ${\rm E}_8\times {\rm E}_8$ model, whereas the second, say
${\scr Z}_-$, produces the ${\rm SO}(16)\times {\rm SO}(16)$ model.
However, the $\bb{Z}_2\times \bb{Z}_2$ twist acts identically in the two
models, and their physical characteristics differ only due to the discrete
torsion eq. \changec.

The projection induced by eqs. \vectortwogamma, or \changec,
has important phenomenological consequences in the free fermionic
constructions that are relevant for orbifold models. In the case of ${\scr
Z}_+$,
the $\bb{Z}_2\times \bb{Z}_2$ orbifold breaks the observable
${\rm E}_8$ symmetry to ${\rm E}_6\times {\rm U}(1)^2$. The chiral matter
states are
contained in the $27$ representation of ${\rm E}_6$, which decomposes as
\eqn\twentysevendecom{
27= 16_{1\over2}+10_{-1}+1_2
}
under its ${\rm SO}(10)\times {\rm U} (1)$ subgroup, where the spinorial $16$
and vectorial $10$  representations of SO(10) contain the Standard Model
fermion
and Higgs states, respectively. The projection induced by \changec\
in ${\scr Z}_-$ entails that either the spinorial  or the vectorial
representation
survives the GSO projection at a given fixed point. Hence, this projection
operates a Higgs-matter splitting mechanism \higgsmattersplit\
in the phenomenological free fermionic models.

Semi-realistic four-dimensional heterotic vacua have also been built using
orbifold technique, based on a choice of gauge bundle and geometrical twist.
These constructions are based essentially on the ${\rm E}_8 \times {\rm E}_8$
heterotic string, and the breaking of ${\rm E}_8$ is achieved by suitable
Wilson lines
\nilles\ (for constructions based on the SO(32) heterotic string see {\it e.g.}
\saul). In this set-ups, different heterotic vacua can be connected by choices
of different gauge bundles and Wilson lines. Although the equivalence of
geometrical $\bb{Z}_2$'s orbifold constructions and free-fermionic
constructions is rather obvious, a explicit link between the two approaches is
still missing and, in particular, to date it is not know how to interpret the
spinor-vector duality in the realm of orbifold compactification. In the next
section, we shall try to fill this gap by analysing a specific $\bb{Z}_2^n$
orbifold and will identify the spinor-vector splitting in terms of discrete
torsion. Connecting the choices of discrete torsion to the choice of gauge
bundles, along the lines of \ploger, is an interesting open problem that we are
able to answer only in the simple case of ${\scr N}=4$ vacua.

%:orbifolds
\newsec{Spinor-vector duality in four-dimensional ${\scr N}=2$ orbifold vacua}

As anticipated, the emergence of spinorial representations in the twisted
sectors of four-dimensional ${\scr N}=1$ heterotic vacua based on $\bb{Z}_2^n$
orbifolds has its origin in the simpler context of vacua with eight
supercharges, where the ${\rm E}_8$ gauge group is directly broken to an
orthogonal one.

To be specific, we consider the ${\rm E}_8 \times {\rm E}_8$ heterotic string
compactified on the $(T^4 \times T^2 )/ \bb{Z}_2 \times \bb{Z}_2^{\prime}
\times \bb{Z}_2^{\prime \prime}$ orbifold.
The factorisation of the internal $T^6$ in terms of the product of a four-torus
times a two-torus is suggested by the way the three $\bb{Z}_2$'s act on the
various degrees of freedom. In particular, the free action generated by
$\bb{Z}_2 \times \bb{Z}_2^\prime$, with
$$
\bb{Z}_2 \ni g = (-1)^{F_1}\, \delta \,,
\qquad \qquad
\bb{Z}_2^\prime \ni g' = (-1)^{F_2}\, \delta \,,
$$
where $F_{1,2}$ flips the sign of the spinorial representation in ${\rm E}_8 =
{\rm Spin} (16)/\bb{Z}_2$ and $\delta$ shifts the compact $x^4$ coordinate by
half of its period,
spontaneously break the ${\rm E}_8 \times {\rm E}_8$ gauge group into ${\rm SO}
(16) \times {\rm SO} (16)$, while preserving the original ${\scr N}=4$
supersymmetries in four-dimensions. The additional $\bb{Z}_2^{\prime\prime}$
factor, instead, twists also the space-time degrees of freedom and preserves
only ${\scr N}=2$ supersymmetries. Its generator $g^{\prime \prime}$ reverts
the sign of the four internal coordinates $x^i$, $i=6,7,8,9$, and, at the same
time, breaks one SO(16) gauge factor (the first one, say) into ${\rm SO}(12)
\times {\rm SO} (4)$.

To implement the action of the $\bb{Z}_2 \times \bb{Z}_2^{\prime} \times
\bb{Z}_2^{\prime \prime}$ orbifold, it is convenient to break the
ten-dimensional SO(8) little group into ${\rm SO} (4) \times {\rm SO} (4)$,
where the second SO(4) factor reflects the symmetry of the internal $T^4$,
while the first SO(4) factor corresponds to the ``enhanced'' little group of
${\scr M}_{1,3} \times T^2$. At the same time, the first ${\rm Spin} (16)$
group factor is broken into ${\rm Spin} (12) \times {\rm Spin} (4)$. As a
result, the one-loop partition function can be written in terms of the familiar
space-time characters, $Q_o$, $Q_v$, $Q_s$ and $Q_c$, and the gauge-group ones,
$\chi^o_i$, $\chi^v_i$, $\chi^s_i$, $\chi^c_i$ and $\xi^{o,v}_{1,g'}$. For
completeness, their explicit expression in terms of ${\rm SO} (2n)$ characters
\openstring\
is given in the appendix. The corresponding genus one partition function thus
reads
$$
{\scr Z} = {1\over 8} \sum_\alpha {\scr Z}_\alpha\,,
$$
where $\alpha$ labels the eight  (un)twisted sectors and each amplitude
${\scr Z}_\alpha$ is given explicitly by
$$
\eqalign{
{\scr Z}_1 =& \Biggl\{ \left(\bar Q_o + \bar Q_v \right) \, \left[
\chi^o_{1} \, \xi^o_{1} + \chi^o_{g} \, \xi^o_{g'} +
(-1)^m\, \left(  \chi^o_{g} \, \xi^o_{1} + \chi^o_{1} \, \xi^o_{g'} \right)
 \right]\, \varLambda^{(4,4)}
\cr
&+ \left(\bar Q_o - \bar Q_v \right) \, \left[
\chi^o_{g''}\, \xi^o_{1} +\chi^o_{g\,g''} \, \xi^o_{g'} +
(-1)^m \, \left( \chi^o_{g\,g''} \, \xi^o_{1} + \chi^o_{g''}\,
\xi^o_{g'}\right) \right]
\,  \left| {2 \eta \over \vartheta_2}\right|^4 \Biggr\}
\cr
&\times \varLambda^{(2,2)} \,,
\cr}
$$
\eqn\zg{
\eqalign{
{\scr Z}_g =& \Biggl\{ \left(\bar Q_o + \bar Q_v \right) \,
\left[  \chi^v_1\, \xi^o_1 -\epsilon_1 \, \chi^v_g \, \xi^o_{g'}
- (-1)^m\, \left( \chi^v_g\, \xi^o_1 -\epsilon_1\, \chi^v_1 \, \xi^o_{g'}
\right)\right]
\, \varLambda^{(4,4)}
\cr
&+ \left( \bar Q_o - \bar Q_v \right) \, \left[
\epsilon_2 \, \chi^v_{g''}\, \xi^o_1 -\epsilon_3\, \chi^v_{g \, g''}\,
\xi^o_{g'}
-(-1)^m \, \left( \epsilon_2\, \chi^v_{g\, g''}\, \xi^o_1 - \epsilon_3\,
\chi^v_{g''}\, \xi^o_{g'} \right) \right] \, \left| {2 \eta \over \vartheta_2
}\right|^4 \Biggr\}
\cr
&\times \varLambda^{(2,2)}_{1/2}\,,
\cr}
}
\eqn\zgp{
\eqalign{
{\scr Z}_{g'} =& \Biggl\{ \left( \bar Q_o + \bar Q_v \right) \, \left[
\chi^o_1\, \xi^v_1 -\epsilon_1 \, \chi^o_g \, \xi^v_{g'} +
(-1)^m \left( \epsilon_1\, \chi^o_g\, \xi^v_1 - \chi^o_1 \, \xi^v_{g'} \right)
\right]
\, \varLambda^{(4,4)}
\cr
&+ \left( \bar Q_o - \bar Q_v \right) \, \left[
\epsilon_4 \, \chi^o_{g''}\, \xi^v_1 -\epsilon_5\, \chi^o_{g\, g''} \,
\xi^v_{g'}
+(-1)^m \left( \epsilon_5 \, \chi^o_{g\, g''}\, \xi^v_1 -\epsilon_4\,
\chi^o_{g''}\, \xi^v_{g'}
\right)\right] \, \left| {2\eta \over \vartheta_2}\right|^4 \Biggr\}
\cr
&\times \varLambda^{(2,2)}_{1/2}\,,
\cr}
}
\eqn\zggp{
\eqalign{
 {\scr Z}_{g\, g'} =& \Biggl\{ \left( \bar Q_o + \bar Q_v \right) \, \left[
\chi^v_1\, \xi^v_1 + \chi^v_g \, \xi^v_{g'}
-\epsilon_1\, (-1)^m\, \left( \chi^v_g\, \xi^v_1 + \chi^v_1 \, \xi^v_{g'}
\right)\right]
\, \varLambda^{(4,4)}
\cr
&+\left( \bar Q_o - \bar Q_v \right) \, \left[
\epsilon_6 \left( \chi^v_{g''}\, \xi^v_1 + \chi^v_{g\, g''}\, \xi^v_{g'}\right)
+ \epsilon_7\, (-1)^m \, \left( \chi^v_{g\, g''}\, \xi^v_1 + \chi^v_{g''}\,
\xi^v_{g'} \right) \right]
\,  \left| {2 \eta \over \vartheta_2 }\right|^4\Biggr\}
\cr
&\times \varLambda^{(2,2)}\,,
\cr}
}
$$
\eqalign{
 {\scr Z}_{g''}=& \Biggl\{ \left( \bar Q_s + \bar Q_c \right) \, \left[
\chi^c_1 \, \xi^o_1 - \epsilon_6\, \chi^c_g \, \xi^o_{g'}
-(-1)^m \, \left( \epsilon_2\, \chi^c_g\, \xi^o_1 - \epsilon_4\, \chi^c_1\,
\xi^o_{g'} \right) \right]
\, \left| {2 \eta\over \vartheta_4}\right|^4
\cr
&+ \left( \bar Q_s - \bar Q_c \right) \, \left[
\chi^c_{g''}\, \xi^o_1 - \epsilon_6\, \chi^c_{g\, g''}\, \xi^o_{g'}
-(-1)^m\, \left( \epsilon_2\, \chi^c_{g\, g''}\, \xi^o_1 -\epsilon_4\,
\chi^c_{g''}\, \xi^o_{g'}
\right)\right] \,  \left| {2 \eta\over \vartheta_3}\right|^4 \Biggr\}
\cr
&\times \varLambda^{(2,2)}\,,
\cr}
$$
$$
\eqalign{
{\scr Z}_{g\, g''}=&\Biggl\{ \left(\bar Q_s + \bar Q_c \right)\, \left[
\chi^s_1\, \xi^o_1 -\epsilon_7\, \chi^s_g\, \xi^o_{g'}
+(-1)^m \, \left( \epsilon_2 \, \chi^s_g\, \xi^o_1+ \epsilon_5\, \chi^s_1\,
\xi^o_{g'} \right) \right]\,
\left| {2\eta \over \vartheta_4}\right|^4
\cr
&+  \left(\bar Q_s - \bar Q_c \right)\,
\left[ \epsilon_2\, \chi^s_{g''}\, \xi^o_1 + \epsilon_5\, \chi^s_{g\, g''}\,
\xi^o_{g'} + (-1)^m\, \left( \chi^s_{g\, g''}\, \xi^o_1 - \epsilon_7 \,
\chi^s_{g''}\, \xi^o_{g'}
\right) \right]\, \left| {2\eta \over \vartheta_3}\right|^4 \Biggr\}
\cr
&\times \varLambda^{(2,2)}_{1/2}\,,
\cr}
$$
$$
\eqalign{
{\scr Z}_{g'\, g''} =& \Biggl\{ \left( \bar Q_s +\bar Q_c \right) \left[
\chi^c_1\, \xi^v_1 - \epsilon_7\, \chi^c_g \, \xi^v_{g'} - (-1)^m \, \left(
\epsilon_3\, \chi^c_g\, \xi^v_1 +\epsilon_4\, \chi^c_1\, \xi^v_{g'} \right)
\right]\, \left| {2\eta \over \vartheta_4}\right|^4
\cr
&+  \left( \bar Q_s - \bar Q_c \right) \left[
\epsilon_4\, \chi^c_{g''}\, \xi^v_1 +\epsilon_3\, \chi^c_{g\, g''}\, \xi^v_{g'}
+(-1)^m\, \left(
\epsilon_7 \, \chi^c_{g\, g''}\, \xi^v_1 - \chi^c_{g''}\, \xi^v_{g'}
\right)\right]\,
\left| {2\eta \over \vartheta_3}\right|^4 \Biggr\}
\cr
&\times \varLambda^{(2,2)}_{1/2}\,,
\cr}
$$
and, finally,
$$
\eqalign{
{\scr Z}_{g\, g'\, g''} =& \Biggl\{\left( \bar Q_s + \bar Q_c \right) \,
\left[
\chi^s_1\, \xi^v_1 -\epsilon_6\, \chi^s_g\, \xi^v_{g'} + (-1)^m \, \left(
\epsilon_3\, \chi^s_g\, \xi^v_1 - \epsilon_5\, \chi^s_1 \, \xi^v_{g'} \right)
\right]\, \left|
{2\eta\over \vartheta_4}\right|^4
\cr
&+ \left( \bar Q_s - \bar Q_c \right) \, \left[
\epsilon_6\, \chi^s_{g''}\, \xi^v_1 - \chi^s_{g\, g''}\, \xi^v_{g'} + (-1)^m\,
\left(
\epsilon_5\, \chi^s_{g\, g''}\, \xi^v_1 - \epsilon_3 \, \chi^s_{g''}\,
\xi^v_{g'} \right) \right]\,
\left| {2\eta \over \vartheta_3}\right|^4 \Biggr\}
\cr
&\times \varLambda^{(2,2)}\,.
\cr}
$$

Before we analyse the properties of the spectrum of this heterotic orbifold, it
is convenient to explain the notation and the origin of the $\epsilon$'s signs.
For convenience, let us take the amplitude ${\scr Z}_g$. It is actually a
short-hand notation for
$$
\eqalign{
{\scr Z}_g =& \int_{\sscr F} {d^2 \tau \over \tau_2^4} \, \sum_{m_4,
m_5, n_4 , n_5} \Biggl[
{\left( \bar Q_o + \bar Q_v \right) \over \bar\eta^2} \, {\left( \chi^v_1 -
(-1)^{m_4} \, \chi^v_g \right) \left( \xi^o_1 +\epsilon_1\, (-1)^{m_4} \,
\xi^o_{g'} \right) \over \eta^2}\,  \varLambda^{(4,4)}
\cr
&+ {\left( \bar Q_o - \bar Q_v \right) \over \bar\eta^2} \, {\left(
\chi^v_{g''} - (-1)^{m_4} \, \chi^v_{g\, g''} \right) \left( \epsilon_2 \,\,
\xi^o_1 + \epsilon_3\,  (-1)^{m_4} \, \xi^o_{g'} \right)  \over \eta^2} \,
 \left| {2 \eta \over \vartheta_2 }\right|^4\Biggr]
 \cr
&\qquad \times \, \varLambda_{m_4, m_5 ; n_4 +{1\over 2},n_5}^{(2,2)} \,,
\cr}
$$
where the eta functions in the denominators count the contribution of the
non-compact world-sheet bosons in the light-cone gauge, while
$$
\varLambda^{(4,4)} = \sum_{m_i , n^i} {q^{{\alpha ' \over 4} p_L^2}\, \bar
q^{{\alpha ' \over 4} p_R^2}
\over \eta^4\, \bar\eta ^4}
$$
denotes the four-dimensional Narain lattice associated to the directions $x^i$,
$i=6,7,8,9$, upon which $g''$ has a non-trivial action. Finally, the (shifted)
zero modes associated to the two remaining compact coordinates fill the lattice
$$
\varLambda_{m_4, m_5 ; n_4 +b ,n_5}^{(2,2)} =
{q^{{\alpha' \over 4} \left( {m_4 \over R_4} + {(n_4 +b) R_4\over \alpha'}
\right)^2 }
\, \bar q^{{\alpha' \over 4} \left( {m_4 \over R_4} - {(n_4 +b) R_4\over
\alpha'} \right)^2}
\over \eta \, \bar \eta} \,
{q^{{\alpha' \over 4} \left( {m_5 \over R_5} + {n_5 R_5\over \alpha'}
\right)^2}
\, \bar q^{{\alpha' \over 4} \left( {m_5 \over R_5} - {n_5 R_5\over \alpha'}
\right)^2}
\over \eta \, \bar \eta}\,,
$$
where $b=0$ in the untwisted, $g''$, $g\, g' $ and $g\, g' \, g''$ twisted
sectors, while $b={1\over 2}$ in the $g$, $g'$, $g\, g''$ and $g'\, g''$
twisted sectors.

The signs $\epsilon_i$ reflect the possibility of turning on discrete torsion
in this $\bb{Z}_2^3$ orbifold.
Clearly, they affect the massless and massive spectrum and in particular the
gauge-group representations of the twisted matter.

This can be neatly seen by writing a $q$-series expansion of the various
contributions to the partition function, and keeping for simplicity only the
low-lying states. In particular, noting that
$$
\eqalign{
q^{n/12}\, V_{2n} &\sim 2n\, q^{1/2} +O(q^{3/2})\,,
\cr
q^{n/12}\, O_{2n} &\sim q^{-1} + n (2n-1)\, +O(q)\,,
\cr}
\qquad
\eqalign{
q^{n/12}\, S_{2n} &\sim 2^{n-1}\, q^{n/2} +O(q^{n/2+1})\,,
\cr
q^{n/12}\, C_{2n} &\sim 2^{n-1}\, q^{n/2}+ O(q^{n/2+1}) \,,
\cr}
$$
and using similar expansions for the theta and eta functions, one finds that
only the untwisted, $g\, g'$, $g''$ and $g\, g'\, g''$ twisted sectors actually
yield massless states. More in details, the leading contributions to the
amplitudes read
$$
{\scr Z}_{(0)} = {\scr Z}_{(0)\, 1} + {\scr Z}_{(0)\, g\, g'} + {\scr Z}_{(0)\,
g''} +{\scr Z}_{(0)\, g\,g'\, g''}\,,
$$
where
$$
{\scr Z}_{(0)\, 1} \sim \bar Q_o \, O_4 \, O_{12}\, O_{16} + \bar Q_v \, V_4 \,
V_{12}\, O_{16}+{\rm massive}\,,
$$
$$
\eqalign{
{\scr Z}_{(0)\, g\, g'} &\sim \bar Q_{o}\,  \left[ O_4 \, V_{12}\, V_{16} \,
{1-\epsilon_1 +\epsilon_6 + \epsilon_7 \over 4} + V_4 \, O_{12}\, V_{16}\,
{1-\epsilon_1 - \epsilon_6 -\epsilon_7 \over 4} \right]
\cr
&+ \bar Q_v \, \left[ O_4\, V_{12}\, V_{16} {1-\epsilon_1 - \epsilon_6
-\epsilon_7 \over 4} + V_4 \, O_{12}\, V_{16} {1-\epsilon_1 +\epsilon_6 +
\epsilon_7 \over 4} \right] + {\rm massive}\,,
\cr}
$$
$$
\eqalign{
{\scr Z}_{(0)\, g''} \sim& 16\, \bar Q_s\, \left[ O_4 \, S_{12} \, O_{16} \,
{1-\epsilon_2 + \epsilon_4 - \epsilon_6 \over 4} + C_4 \, V_{12}\, O_{16}\,
{1+\epsilon_2 + \epsilon_4 + \epsilon_6 \over 4} \right]
\cr
&+ 16\, \bar Q_s\, \left[ S_4 \, O_{12} \, O_{16} \,
{1+\epsilon_2 + \epsilon_4 + \epsilon_6 \over 4}  \right]
+{\rm massive}\,,
\cr}
$$
and, finally,
$$
{\scr Z}_{(0)\, g\,g'\, g''} \sim 16\, \bar Q _s \, C_4 \, O_{12} \, V_{16} \,
{1-\epsilon_3 - \epsilon_5 + \epsilon_6\over 4} + {\rm massive}\,.
$$

The untwisted sector, independent of the discrete torsion, comprises an ${\scr
N}=2$ supergravity multiplet, coupled to vector multiplets in the adjoint
representation of the gauge group $G={\rm U} (1)^2\times{\rm SO} (4) \times
{\rm SO} (12) \times {\rm SO} (16)$, and hypermultiplets in the representation
$(4,12,1)$.

The twisted matter includes neutral hypermultiplets associated to the
deformations of $K3$, together with hypermultiplets charged with respect to
$G$, whose representations depend on the choice of the discrete torsions.
Clearly, for the partition function to be real the $\epsilon$'s can only be
signs, while demanding that ${\scr Z}\,$ have a physical interpretation in
terms of  a proper counting of states, the various coefficients of the $q^\alpha$ terms
must be integers,  positive for bosons and negative for fermions.
This clearly implies that the combinations of discrete
torsion, like those appearing in ${\scr Z}_{(0)}$ should equal 0 or 1.
Finally, the last requirement we want to impose on the $\epsilon$'s is that
the gauge group be the smallest one. In fact, if any
of the combinations in the first line of ${\scr Z}_{(0), g\, g'}$ is different
than zero, the gauge group is enhanced to ${\rm SO} (16) \times
{\rm SO} (16)$ or to ${\rm SO}(12) \times {\rm SO} (20)$. As we shall see
momentarily, this possibility is already present at the level of ${\scr N}=4$
vacua, and corresponds to different discrete values of Wilson lines. Taking all
these constraints into account, the possible choices of discrete torsion turn
out to be
$$
\epsilon_1 =1\,, \quad \epsilon_7 = - \epsilon_6\,, \quad \epsilon_4=\epsilon_5\,,
$$
and
$$
\eqalign{
sol_1 &= (-1,-1,+1,-1)\,,
\cr
sol_2 &=(+1,+1,-1,-1)\,,
\cr}
\qquad
\eqalign{
sol_3 &=(+1,+1,+1,+1)\,,
\cr
sol_4 &=(-1,-1,-1,+1)\,,
\cr}
$$
where
$$
sol_i = ( \epsilon_2 , \epsilon_3 , \epsilon_4 , \epsilon_6) \,.
$$

As a result, the massless twisted spectra depend on the allowed combination
$sol_i$
of signs, and are listed in table 1. This is a neat instance of spinor-vector
duality and is at the heart of Higgs-matter splitting in more realistic vacua.
Let us note that for  $sol_3$ extra $8\times 8$ neutral massless hypermultiplets appear from the twisted sector
$$
\bar Q_s\,  S_4 \, O_{12} \, O_{16} \,,
$$
hence keeping a total number of massless degrees of freedom unchanged.
From table 1 it is observed that under
the different possibilities of the discrete torsions
the number of massless degrees of freedom is preserved, except  $sol_2$ that does not have any twisted messless states, similar
to what is observed in the free fermionic classification of \spinvecduality.
Let us note that there are no other massless twisted neutral hypermultiplets in any of
$sol_i$.

To further break supersymmetry, and get more realistic chiral models, it is
enough to act with an additional $\bb{Z}_2^{\prime\prime\prime}$ that twists
the coordinates $x^{4,5,6,7}$, say, and breaks the ${\rm SO}(12)$ gauge group
to the more phenomenological ${\rm SO} (10)$, while leaving untouched the
hidden sector. Although, additional discrete torsions can be turned on for this
$\bb{Z}_2^4$ model, in the simplest instance where one considers only the seven
signs previously introduced, the resulting massless chiral spectrum for the
choice $sol_1$  includes an SO(10) spinorial representation since,
under the action of $\bb{Z}_2^{\prime\prime\prime}$, $32 \to
16 + \overline{16}$, and only one spinorial eventually survives the overall
orbifold projection in a chiral model. On the other hand, the solutions $sol_3$
and $sol_4$ would only include matter in vectorial (Higgs-like)
representation.

Furthermore, the reduction of the number of chiral families, or alternatively
the change of the topology of the Calabi-Yau manifold, can be achieved, as
usual, through the implementation of additional shift symmetries. They do not
twist any internal coordinate and the only effect on the spectrum consists in
reducing the
number of families of twisted chiral matter through an identification of fixed
points.
In terms of free fermionic constructions, this is equivalent to the inclusion
of the $\{ \alpha ,\beta , \gamma \}$ system of boundary condition basis
vectors to the NAHE set, as discussed in the previous section.

\vskip 30pt
\vbox{
\settabs=6\columns
\centerline{\vbox{\hrule width  12 truecm}}
\+
&solution & $\quad$reps of massless charged hypermultiplets
\cr
\centerline{\vbox{\hrule width  12 truecm}}
\+
&\hskip .4truecm $sol_1$ & \hskip 2 truecm $8\times (1,32,1)$
\cr
\+
&\hskip .4truecm $sol_2$ & \hskip 2 truecm  ---
\cr
\+
&\hskip .4truecm $sol_3$ & \hskip 2 truecm $8\times \left[ (2,12,1) + \, 4 \times (2,1,1)\right]$
\cr
\+
&\hskip .4truecm $sol_4$ & \hskip 2 truecm $8\times (2,1,16)$
\cr
\centerline{\vbox{\hrule width  12 truecm}}
\cleartabs
\vskip 10pt
\noindent
{\ninepoint {\bf Table 1.} Charged massless twisted spectrum for the
$T^6/\bb{Z}_2^3$ heterotic orbifold, for different choices of discrete torsion.
The non-Abelian gauge group is $G={\rm SO} (4) \times {\rm SO} (12) \times {\rm
SO} (16)$ and the vacuum configurations also include universal (untwisted)
charged hypermultiplets in the representation $(4,12,1)$.}
\vskip 30pt
}

Before we conclude, let us make a brief remark on the interpretation of the
$\bb{Z}_2 \times \bb{Z}_2 '$ freely-acting orbifold of the ${\rm E}_8 \times
{\rm E}_8$ heterotic string.  As already stated several times, this orbifold
projection does
not break any of the original supersymmetries,  and therefore corresponds to a
nine-dimensional vacuum with 16 supercharges. Depending on the value of the
discrete torsion\footnote{$^\dagger$}{The discrete torsion present in this
$\bb{Z}_2 \times \bb{Z}_2 '$ actually corresponds to the sign $\epsilon_1$ in
eqs. \zg , \zgp\ and \zggp.} one gets the models
$$
{\scr Z}_{\epsilon_1 = +1} = (\bar V_8 - \bar S_8 ) \, \left[ O_{32}\,
\varLambda_{2m,n} + S_{32}\, \varLambda_{2m+1,n}
+ V_{32}\, \varLambda_{2m+1,n+1/2} +C_{32}\, \varLambda_{2m,n+1/2} \right]\,,
$$
with an SO(32) gauge group, and
$$
\eqalign{
{\scr Z}_{\epsilon_1 =-1} &=  (\bar V_8 - \bar S_8 ) \, \bigl[ (O_{16} \,
O_{16} + C_{16}\, C_{16} ) \, \varLambda_{2m,n}
+ (S_{16} \, S_{16} + V_{16}\, V_{16} ) \, \varLambda_{2m+1,n}
\cr
&+ (C_{16} \, O_{16} + O_{16}\, C_{16} ) \, \varLambda_{2m,n+1/2}
+ (V_{16} \, S_{16} + S_{16}\, V_{16} ) \, \varLambda_{2m+1,n+1/2}  \bigr]\,,
\cr}
$$
with a broken ${\rm SO} (16) \times {\rm SO} (16)$ gauge group.

However, ${\scr N}=4$ vacua are characterised by a
moduli space,  uniquely fixed by its dimension, and therefore by the dimension
of the compactification torus and by the rank of the gauge group. Indeed, the
heterotic vacua obtained as an $S^1$ and $S^1 / \bb{Z}_2 \times \bb{Z}_2 '$
compactification, with or without discrete torsion, are all continuously
connected.
In this respect, the $\epsilon_1$ discrete torsion has a natural geometrical
description in terms of
discrete values of otherwise continuous Wilson lines along the compact $S^1$.
It is tempting to interpret also the remaining signs as specific choices of
gauge bundles and/or Wilson lines as in \ploger. Although this connection seems
quite natural, it is less evident than in the ${\scr N}=4$ case and requires
further analysis.

%:Conclusions
\newsec{Conclusions}

Heterotic string theory is unique among the perturbative string
constructions since it gives rise naturally to the GUT embedding of the
Standard Model matter states in SO(10) and ${\rm E}_6$ representations in a
perturbative, and thus calculable, set-up.
Grand unification is well supported by the pattern of observed fermion
and gauge
boson charges. In the framework of SO(10) gauge theory all the matter states of
a single generation
are unified in the 16
spinorial representation and, a priori, one needs only two
types of representations, the spinorial 16 and the
vectorial 10 representations,
to embed the Standard Model matter and Higgs spectrum.
The framework of ${\rm E}_6$ grand unification has the further property of
incorporating the 16 matter and 10 Higgs states into the
27 representation of ${\rm E}_6$.

As the observed gauge symmetry at low energies consists solely of the
Standard Model one, its embedding into a grand unification
group necessitates that the larger GUT symmetry be broken.
Moreover, grand unification introduces additional difficulties with proton
decay and
neurtino masses. The GUT  symmetry breaking and the
miscellanea issues typically require the introduction of
large representations, like the 126 of SO(10), or the 351 of ${\rm E}_6$.

By producing the gauge and matter structures that
arise in Grand Unified Theories, heterotic string theories
offer new possibilities to tame the problems that arise in field theory GUTs.
To this end, to understand the various alternatives offered by string theory,
it is important to construct quasi-realistic string models and
investigate their properties in detail. The main approaches to this program
are free-fermionic \fff\ and bosonic \refs{\narain,\orbifolds} constructions,
as well as interacting \refs{\gepner,\chang} world-sheet conformal field
theories. The heterotic-string models in their free fermionic formulation
\refs{\fsufive{--}\cfs},
first constructed over two decades ago, are among the most realistic string
vacua constructed to date, though, in recent years comparable
quasi-realistic models have also been constructed using free world-sheet bosons
\refs{\nilles,\saul}. It should be stressed,
however, that the two formulations are closely related and that the
corresponding
string vacua can be described equivalently using both approaches.
Therefore it is natural to assume that for every string model constructed using
free world-sheet fermions
an identical vacuum exists constructed using
free world-sheet bosons. Indeed, the free fermionic models
correspond to $\bb{Z}_2^n$ toroidal orbifolds
at special points in the moduli space.

While the free-fermionic approach can be straightforwardly implemented as
an algebraic set of conditions that facilitate the scan of phenomenological
properties, the free boson approach is more readily adaptable to explore the
underlying moduli dynamics. It is therefore compelling to develop a dictionary
between the two languages. Although the equivalence is anticipated, writing a
detailed
dictionary is often non-trivial. In this paper we investigated this
aspect in some detail. An important feature in the quasi-realistic
free fermionic models is the breaking of the ${\rm E}_8\times {\rm E}_8$
symmetry to
${\rm SO} (16)\times {\rm SO} (16)$ at the level of the underlying ${\scr N}=4$
toroidal
compactification. This breaking is realised in the bosonic construction
in terms of freely acting orbifolds, or alternatively in terms of Wilson
lines. Matter states arise in the free fermion models from $\bb{Z}_2$ twisted
sectors, which break the ${\scr N}=4$ space-time supersymmetry to ${\scr N}=2$.
The next step in building the dictionary between the two classes of
models is therefore to add a $\bb{Z}_2$ orbifold to the two freely acting
orbifolds of \parti.
However, it turns out that the construction is not straightforward.
In the free fermionic models the one-loop partition functions are
generated in terms of boundary condition basis vectors and
GGSO phases. One then builds a space of modular invariant
partition functions, with differing physical characteristics.
In the bosonic representation, on the other hand, a variety of vacua arise
from the freedom to chose the background fields and from the
existence of disconnected modular orbits. The detailed correspondence
between the two representations, while formally well understood and
established, is nevertheless non-trivial and often obscure.
In this paper we addressed this issue with respect to the
twisted matter states and spinor-vector duality, first observed in the
classification of free fermionic models \refs{\fknr,\fkr,\spinvecduality}.
The spinor-vector duality is a property over the full space of
vacua generated by the given set of basis vectors \spinvecduality\ and
corresponds to maps between different choices of GGSO projection
coefficients. In the orbifold language, as demonstrated here, it corresponds
to different choices of discrete torsions, thus extending the map of
\vafawitten.
Two issues are of interest here. The first is
to improve the understanding of the detailed correspondence between
the free fermion GGSO projection coefficients and the orbifold
discrete torsions. The second is to understand the
spinor-vector duality in geometrical terms. We anticipate that this
should entail an action on the internal moduli plus an action on
the bundle that generates the heterotic-string gauge degrees of freedom.
We note that the existence of the spinor--vector duality raises basic
questions in regard to the relation of the string vacua to the 
low energy effective field theory. While in the effective field theory
the two models identified under the spinor--vector duality map
are clearly distinct, from the string point of view they are 
closely related. This is exemplified by the fact that the number
of degrees of freedom is preserved under the map. Thus, whereas
the spinor of SO(12) contains 32 states and the 
vector contains only $2\times 12=24$, the vector representation
is augmented by additional 8 SO(12) singlets, that correct
the mismatch.
The issue can be further explored by studying
the effect of the duality map on interactions.
Another issue of further interest is the relation of the
spinor--vector duality to triality of SO(8). This 
question was briefly explored in the context of the 
free fermionic classification \spinvecduality\ by breaking
the untwisted gauge degrees of freedom to four SO(8) factors.
We hope to address these issues in future publications.

 \vskip 1.5truecm

%:Acknowledgement
{\bf Acknowledgement.} It is a pleasure to thank Elisa Manno, Cristina
Timirgaziu and Michele Trapletti for stimulating discussions.
CA would like to thank CERN PH-TH and the Department of
Mathematical Sciences, Liverpool, for their kind hospitality during various
stages of this work. MT is grateful to the Department of Theoretical Physics of
The University of Turin for the kind hospitality extended to him during the
final stage of the project.
AEF also thanks the \'Ecole Normale Sup\'eriuere in Paris for hospitality and
the Royal Society for support.
CA is supported in part by the Italian MIUR-PRIN contract 20075ATT78 and in
part by
the ERC Advanced Grant no. 226455, ÒSupersymmetry, Quantum Gravity and
Gauge FieldsÓ (SUPERFIELDS). AEF and MT are supported by a STFC rolling 
grant ST/G00062X/1.

%:appendix
\appendix{A}{List of characters}

In this appendix, we list the space-time and gauge-group characters that enter
in the $T^6/\bb{Z}_2^3$ partition function. For the definition of the ${\rm SO}
(2n)$ characters in terms of eta and theta functions, as well as for their
modular properties, we report the interested reader to \openstring.

Space-time (anti-holomorphic) characters
$$
\eqalign{
\bar Q_o &= \bar V_4 \, \bar O_4 - \bar C_4 \, \bar C_4 \,,
\cr
\bar Q_s &= \bar O_4 \, \bar C_4 - \bar S_4 \, \bar O_4 \,,
\cr}
\qquad \quad
\eqalign{
\bar Q_v &= \bar O_4 \, \bar V_4 - \bar S_4 \, \bar S_4 \,,
\cr
\bar Q_c &= \bar V_4 \, \bar S_4 - \bar C_4 \, \bar V_4 \,.
\cr}
$$

Gauge-group (holomorphic) characters associated to the first ${\rm E}_8 \to
{\rm SO} (4)\times {\rm SO} (12)$ factor
$$
\eqalign{
\chi^o_1 &= O_4 \, O_{12} + V_4\, V_{12} + S_4 \, S_{12} + C_4 \, C_{12}\,,
\cr
\chi^o_g &= O_4 \, O_{12} + V_4\, V_{12} - S_4 \, S_{12} - C_4 \, C_{12}\,,
\cr
\chi^o_{g''} &= O_4 \, O_{12} - V_4\, V_{12} - S_4 \, S_{12} + C_4 \, C_{12}\,,
\cr
\chi^o_{g\, g''} &= O_4 \, O_{12} - V_4\, V_{12} + S_4 \, S_{12} - C_4 \,
C_{12}\,,
\cr}
$$
$$
\eqalign{
\chi^v_1 &= O_4 \, V_{12} + V_4\, O_{12} + S_4 \, C_{12} + C_4 \, S_{12}\,,
\cr
\chi^v_g &= O_4 \, V_{12} + V_4\, O_{12} - S_4 \, C_{12} - C_4 \, S_{12}\,,
\cr
\chi^v_{g''} &= O_4 \, V_{12} - V_4\, O_{12} - S_4 \, C_{12} + C_4 \, S_{12}\,,
\cr
\chi^v_{g\, g''}&= O_4 \, V_{12} - V_4\, O_{12} + S_4 \, C_{12} - C_4 \,
S_{12}\,,
\cr}
$$
$$
\eqalign{
\chi^c_1 &= O_4 \, S_{12} + V_4\, C_{12} + S_4 \, O_{12} + C_4 \, V_{12}\,,
\cr
\chi^c_g &= O_4 \, S_{12} + V_4\, C_{12} - S_4 \, O_{12} - C_4 \, V_{12}\,,
\cr
\chi^c_{g''} &= O_4 \, S_{12} - V_4\, C_{12} - S_4 \, O_{12} + C_4 \, V_{12}\,,
\cr
\chi^c_{g\, g''}&= O_4 \, S_{12} - V_4\, C_{12} + S_4 \, O_{12} - C_4 \,
V_{12}\,,
\cr}
$$
and
$$
\eqalign{
\chi^s_1 &= O_4 \, C_{12} + V_4\, S_{12} + S_4 \, V_{12} + C_4 \, O_{12}\,,
\cr
\chi^s_g &= O_4 \, C_{12} + V_4\, S_{12} - S_4 \, V_{12} - C_4 \, O_{12}\,,
\cr
\chi^s_{g''} &= O_4 \, C_{12} - V_4\, S_{12} - S_4 \, V_{12} + C_4 \, O_{12}\,,
\cr
\chi^s_{g\, g''}&= O_4 \, C_{12} - V_4\, S_{12} + S_4 \, V_{12} - C_4 \,
O_{12}\,.
\cr}
$$
Gauge-group (holomorphic) characters associated to the second  ${\rm E}_8 \to
{\rm SO} (16)$ factor
$$
\eqalign{
\xi^o_1 &= O_{16} + S_{16}\,,
\cr
\xi^o_{g'} &= O_{16} - S_{16}\,,
\cr}
\qquad\quad
\eqalign{
\xi^v_1 &= V_{16} + C_{16} \,,
\cr
\xi^v_{g'} &= V_{16} - C_{16}\,.
\cr}
$$
Notice that the characters $\xi^{o,v}_{1,g'}$ are exactly equal to
$\chi^{o,v}_{1,g}$. We use a different notation only to stress that on the
second ${\rm E}_8$  only the $g'$ generator acts non-trivially and therefore
the group is simply broken to ${\rm SO} (16)$.

%:References
\listrefs

\end